\documentclass{article}
\usepackage [utf8] {inputenc}
\usepackage [english]{babel}
\usepackage [authoryear, longnamesfirst]{natbib}
\usepackage{hyperref}
\usepackage {graphicx}
\usepackage {amsfonts}
\usepackage {amsthm}
\usepackage {amsmath}

\usepackage[dvipsnames]{xcolor}

\usepackage{color}

\title{Modeling Oral Glucose Tolerance Test (OGTT) data and its Bayesian Inverse Problem}
\author{Nicolás E. Kuschinski$^a$, J. Andrés Christen$^a$,\\ Adriana Monroy$^b$, Silvestre Alavez$^c$\\
$^a$\textit{CIMAT, Guanajuato, Gto., Mexico}\\
$^b$\textit{Hospital General, Mexico City, Mexico}\\
$^c$\textit{Universidad Autónoma Metropolitana, Unidad Lerma, Mexico City, Mexico}}
\date{}
\begin{document}
\maketitle{}
\begin{abstract}
    One common way to test for diabetes is the Oral Glucose Tolerance Test or OGTT. Most common methods for the analysis of the data on this test are wasteful of much of the information contained therein. We propose to model blood glucose during an OGTT using a compartmental dynamic model with a system of ODEs.   Our model works well in describing most scenarios that occur during an OGTT considering only 4 parameters.  Fitting the model to data is an inverse problem, which is suitable for Bayesian inference. Priors are specified and posterior inference results are shown using real data.
\end{abstract}

\section{Introduction}
Type 2 diabetes is a serious and common illness. It typically develops without obvious symptoms, and can go undiagnosed for years. With timely diagnosis and treatment type 2 diabetes can change from a serious and potentially life threatening illness to a relatively mild condition. Similarly, a patient who is informed that he or she is at risk of developing type 2 diabetes can often take appropriate measures and thus prevent diabetes from occurring altogether \citep{prevalence, diabetesinfo}.

Diabetes occurs when the body is unable to adequately regulate blood glucose. The body's main mechanism for reducing blood glucose is insulin, which is a hormone produced by the pancreas in response to high levels of blood sugar. Diabetes occurs either because insulin production is insufficient, or because the insulin being produced is ineffective. In either case, the result is that the body cannot reduce blood glucose to healthy levels \citep{diabetesinfo}.

For diagnosis of type 2 diabetes, one common test is the Oral Glucose Tolerance Test, or OGTT.  For this test, a fasting patient arrives and his or her resting glucose is measured from a blood sample. The patient then drinks a 75g glucose concentrate and blood glucose is measured repeatedly over the course of the next two hours.  The exact glucose measuring times vary depending on local practices. The results of these measurements are expected to provide some notion of how the patient's body handles the glucose \citep{prev, ogtt,glucose}.

In practice, the analysis of OGTT tests is usually done using very simple guidelines. Typically used markers include the average of the observed glucose measurements and/or the value of the first and last measurement.  A patient is considered diabetic if the measurement chosen is above a certain threshold (typically $200mg/dl$).  While this analysis has proven to be useful, it disregards one of the primary qualities of OGTT test conditions: That they measure the \textit{evolution} of a process over time \citep{ogtt}.

Accordingly, here we propose a dynamic model based on Ordinary Differential Equations (ODEs) to model blood glucose during an OGTT.  The idea of using mathematical models to analyze OGTT results is not new.  Previously proposed models have not been used for inference, mostly because they lack the flexibility to explain many of the phenomena seen in OGTTs. For instance, \cite{prev} assumes that the body only ever lowers blood glucose, but in the course of measuring several patients we can see that this is not always the case (see real data in section \ref{results})

In our approach, we use a dynamic model which was derived from the recommendations of our medical collaborators and follows the logic of previous related works such as \cite{othermodel}. It is flexible enough to describe most of the observed behavior of glucose in real patient's OGTTs. Using Bayesian inference, we put appropriate priors on the parameters and fit this model to real data. We have been able to achieve good fits for observed data and our results match the intuition of our medical collaborators well enough that we consider it a good candidate to be considered for serious analysis of OGTT data and eventually for early diasgnosis.

The paper is organized as follows, in section \ref{sec:dyn} we present the dynamic model. In section \ref{sec:stat} We develop a Bayesian statistical model that can be used to draw inference from the dynamic model. In section \ref{sec:inference} we explain the details of how to perform inference from the model, and present results of said inference on real patients. Finally, section \ref{conclusions} concludes the paper.

\section{The dynamic model}\label{sec:dyn}

Our model is based on the interaction of glucose, insulin and glucagon only.  The glucose regulation system is far more complex but in the controlled environment of an OGTT these are by far the leading factors. Insulin is a hormone secreted by the pancreas which reduces blood glucose. Glucagon is also a hormone produced in the pancreas and has the opposite effect, it triggers the liver to produce glucose, thus increasing blood glucose levels.  In simple terms, insulin is produced when blood glucose is high and glucagon is produced when blood glucose is low, to make a feedback system of blood glucose level regulation \citep{glucagon, othermodel}.

Our dynamical model is represented by the following system of ODEs
\begin{align}
    \frac{dG}{dt} &= L-I+\frac{D}{\theta_2}\label{eq:dyn1}\\
    \frac{dI}{dt} &= \theta_0(G-G_b)^+-\frac{I}{a}\label{eq:dyn2}\\
    \frac{dL}{dt} &= \theta_1(G_b-G)^+-\frac{L}{b}\label{eq:dyn3}\\
    \frac{dD}{dt} &= -\frac{D}{\theta_2}+\frac{2V}{c}\label{eq:dyn4}\\
    \frac{dV}{dt} &= -\frac{2V}{c}\label{eq:dyn5}
\end{align}
where the meaning of each  of the state variables and parameters is explained in table \ref{tab.variables}.

\begin{table}
        \begin{tabular}{| c | l | c |} \hline
            &  Interpretation & Value \\ \hline
            $G$ & Blood glucose. & State variable\\ \hline
            $I$ & Blood Insulin.  & State variable\\ \hline
            $L$ & Blood Glucagon.  & State variable\\ \hline
            $D$ & Glucose in digestive system.  & State variable\\ \hline
            $V$ & Glucose not yet in the digestive system & State variable\\ \hline
            $\theta_0$ & Insulin responsiveness  & Unknown par.\\ \hline
            $\theta_1$ & Glucagon responsiveness  & Unknown par.\\ \hline
            $\theta_2$ & Glucose digestive system mean life.  & Unknown par.\\ \hline
            $a$, $b$ & Insulin and Glucagon clearance mean life. & 31 min. \\ \hline 
            $c$ & Time taken to drink most of the glucose solution & 5 min max. \\
            \hline
        \end{tabular}  
        \caption{\label{tab.variables} Meanings of state variables and parameters in the OGTT model.}
\end{table}

The heuristics behind this model are similar to other glucose-insulin models \citep[for instance]{othermodel} and are as follows.  There is a threshold level of glucose which the body hopes to maintain which is denoted by $G_b$. It is set at $80mg/dl$ for all examples in this paper, but it can be adjusted or inferred otherwise if that is deemed appropriate. If blood glucose goes above $G_b$ then insulin is produced, increasing $\frac{dI}{dt}$ as indicated by (\ref{eq:dyn2}). As insulin is produced, this acts to reduce glucose concentration in the blood, reducing $\frac{dG}{dt}$ as indicated by (\ref{eq:dyn1}). The opposite effect is achieved by glucagon, as seen in (\ref{eq:dyn3}) and (\ref{eq:dyn1}).  Insulin and glucagon are both metabolized and decrease with mean lifes $a$ and $b$ as seen in equations~(\ref{eq:dyn2}) and~(\ref{eq:dyn3}), respectively.

$D(t)$ and $V(t)$ represent glucose which is moving into the bloodstream. It begins outside the body, ie. the sugar concentrate $V(t)$, decreasing and moving into the digestive system, $D(t)$, as seen in (\ref{eq:dyn5}) and (\ref{eq:dyn4}), and then from the digestive system moving into the bloodstream, as seen in (\ref{eq:dyn4}) and~(\ref{eq:dyn1}).

Time is measured in hours, and blood glucose is measured in $mg/dL$ of blood. The units of glucagon and insulin are more abstract and can be thought of in terms of their effect on units of blood glucose. Insulin and glucagon responsiveness include both the generation of the hormone and also the response of the body to the hormone after production.  The model is not intended for insulin nor glucagon level prediction and only glucose measurements are available, therefore in our model the units of $I$ and $L$ are not relevant and not directly interpretable.

$a$ and $b$ are extrapolated from best estimates of insulin and glucagon clearance time from \cite{insulindeg}. Similarly, estimates exist on times of glucose absorption into the body \citep{glucose}, but these vary greatly from patient to patient and thus $\theta_2$ is inferred.  Jointly with $\theta_0$ and $\theta_1$, which are also inferred, these parameters represent the patient's condition in our model.

For the examples in this paper, the system of ODEs is solved numerically (there is no known analytic solution). This is done by using the odeint function in the scipy package of the python programming language, \citep{python}. This uses an implementation of the LSODA algorithm, described in \cite{lsoda}.

\begin{figure}
    \begin{tabular}{c c}
        \includegraphics[scale=0.3]{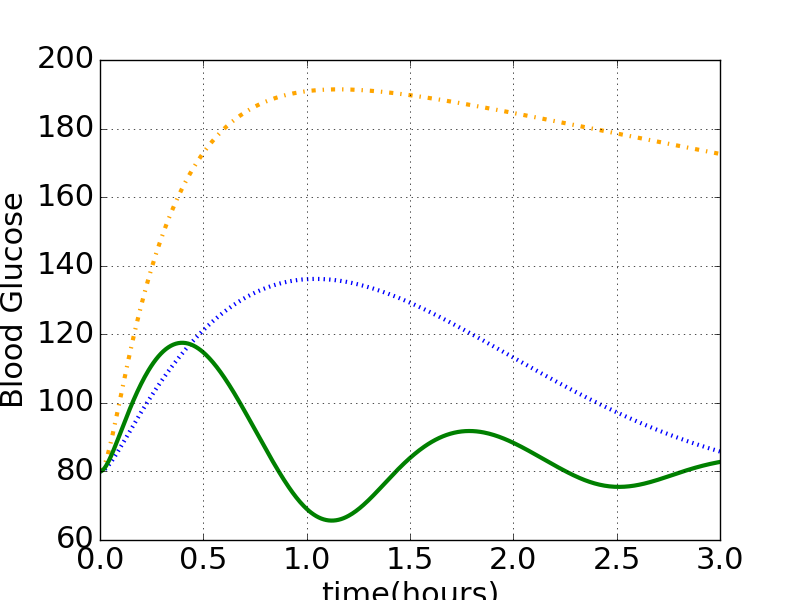}&
        \includegraphics[scale=0.3]{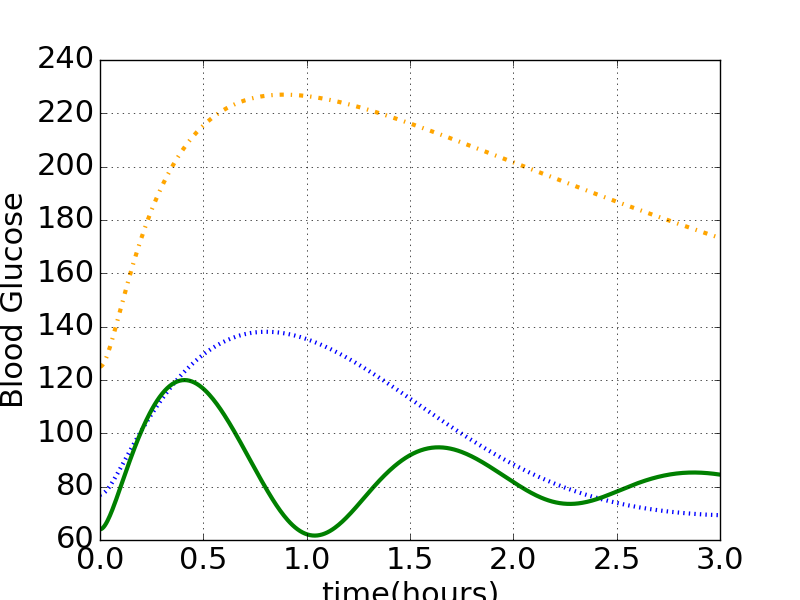}
    \end{tabular}
    \caption{\textbf{Left:} Three curves produced by our model, all beginning with $G(0)=G_b=80mg/dl$ these represent three kinds of patient: The dotted line is a healthy normal patient, the broken line is a diabetic patient who does not adequately regulate insulin, and the solid line is an oscillating patient, whose insulin and glucagon response is very strong. \textbf{ Right:}  Curves showing similar scenarios but with slightly different parameter values, including $G(0)$.}
    \label{fig:curves}
\end{figure}

While some justification for the dynamic model comes from the heuristics, this is secondary to the real issue, which is whether its behavior can adequately represent what happens to glucose inside a patient's body. In figure \ref{fig:curves} we see glucose curves which follow from the model. The first three curves all start at $G_b$, to show the behavior of the model when the patient is already stable. One of the purposes of asking patients to fast beforehand is precisely to obtain this behavior -- however, particularly for diabetic patients, fasting may not be sufficient and glucose may begin elsewhere. The curves in the right panel of figure \ref{fig:curves} represent a scenario wherein glucose begins somewhere other than $G_b$.

\section{Statistical model for the OGTT data analysis: The Inverse Problem} \label{sec:stat}

In order to perform inference on the OGTTs of real patients, the model must be fit to the data, ie. the patient's glucose readings over the course of the test.  For instance, for one real patient, at times $t=$ 0:00,\, 0:30,\, 1:00,\, 1:30, and 2:00 hours we obtained glucose measurements of $y=81,156,141,102$, and $89$ $mg/dl$ respectively. The intent is to use these data to infer the glucose curves. We assume data to be observations of $G(t)$ at the measured times and model the data $y$ with
$$
y_i=G(t_i)+\epsilon_i
$$
where $\epsilon_i\sim\mathcal{N}(0,\sigma^2)$ and $\sigma=5mg/dl$ for all examples in this paper. This allows us to write the likelihood as
$$
f(y|\theta)\propto\prod_ie^{-\frac{y_i-G(t_i|\theta)}{2\sigma^2}}.
$$

Fitting this kind of model is considered an inverse problem; a non-linear regression problem with a complex regressor defined through a system of ODEs.  These systems are frequently characterized by drastically different sets of parameters fitting well with the same data, leading to many explanatory scenarios. For this reason, classical statistical estimators, such as the least squares estimator, only select one possible scenario, often quite an unreasonable one, following the data closely. There are several ways to address this issue, but one popular choice is to use Bayesian inference and encode some notion of what reasonable parameter combinations are in the prior distribution \citep{inverse, inversebook}.

For all of the examples in this paper, the following priors were used:
\begin{align*}
    \theta_0&\sim Gamma(2,0.25)\\
    \theta_1&\sim Gamma(2,0.25)\\
    \theta_2&\sim Gamma(10,20)\\
    G_b&\sim \mathcal{N}(80,20000)~ \text{truncated to} ~[30,400].
\end{align*}

The priors for $\theta_0$ and $\theta_1$ were chosen to give high probability to all values estimated from even the most extreme patients that have been analyzed in this way. The prior for $\theta_2$ is chosen to match information in \cite{glucose}. The prior for $G_b$ is centered on healthy patients and is truncated since any patient whose initial glucose is outside of this range should not undergo an OGTT test but instead be placed into emergency care (AM performs a preliminary finger stick glucose test precisely for this purpose).

\section{Inference}\label{sec:inference}

The object of interest is the function $G(t)$ for each patient and inference is performed on data from each patient separately leading to a separate posterior for $\theta_0, \theta_1, \theta_2$ and $G_b$ for each patient. Our objective here is not a population study, and hence we  concentrate on studying our model and its ability to fit OGTT data parsimoniously.  Posterior exploration is achieved using MCMC techniques. Most MCMCs must be tuned to the posterior for each situation and in this case for each patient. A practical alternative is to use a self-tuning MCMC algorithm. One such algorithm is the t-walk, which is an MCMC algorithm that adapts to the scale of the target distribution. This is the algorithm that was chosen for this case, see \cite{twalk}.

Posterior exploration can be done in reasonable time even without high end hardware. All the examples in this paper were performed on a laptop computer with an i5 processor and took less than 2 minutes to perform 15000 iterations of the t-walk.  This represents 150 pseudo-independent posterior samples (using higher than necessary autocorrelation times, to account for patients with posterior distributions which are harder to explore than usual). This is quite an acceptable numerical processing time, since it takes 2 hours to gather the blood samples and processing is typically done overnight, depending on the availability of staff and laboratory equipment.

\subsection{Results}\label{results}

\begin{figure}\begin{tabular}{c}
    \includegraphics[scale=0.3]{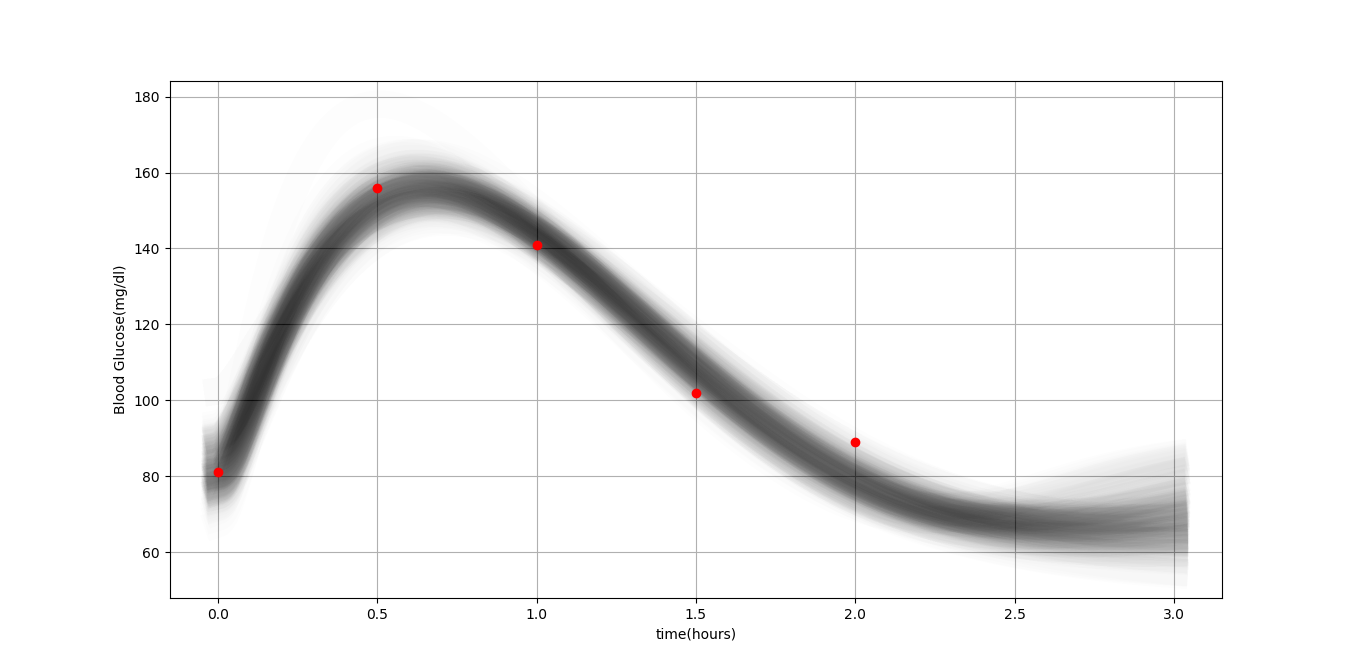} \\
    \includegraphics[scale=0.3]{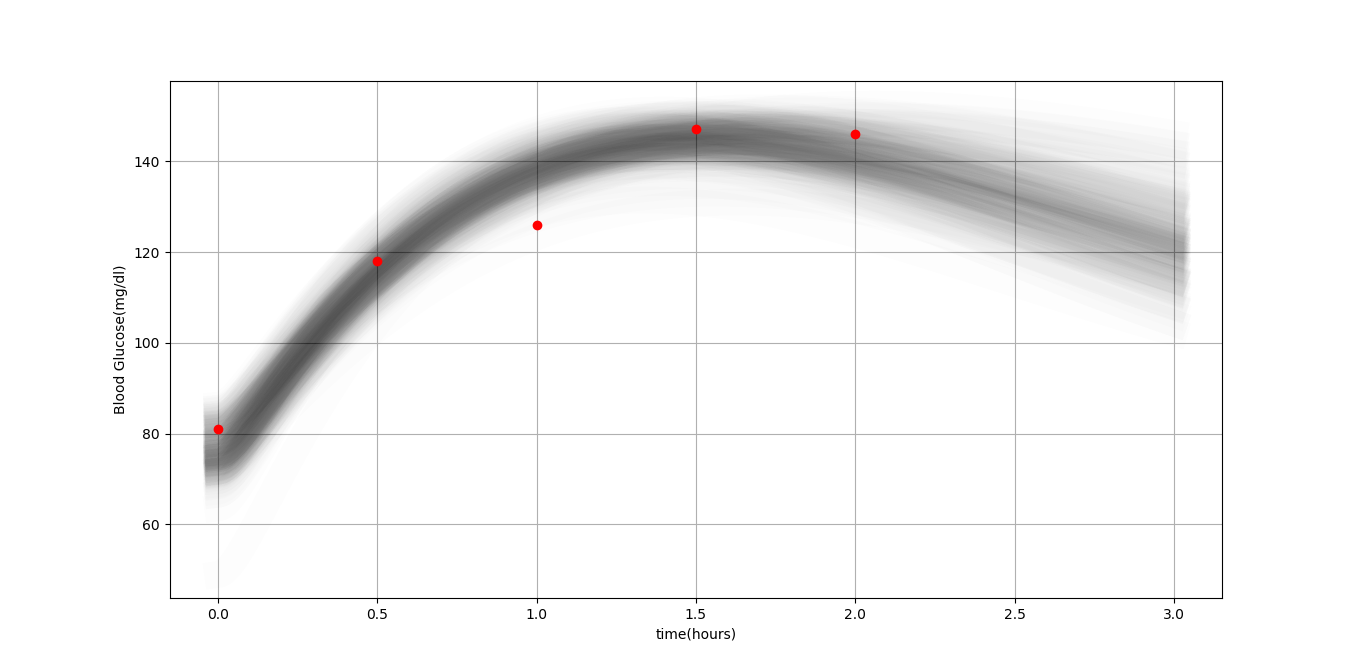} \\
    \includegraphics[scale=0.3]{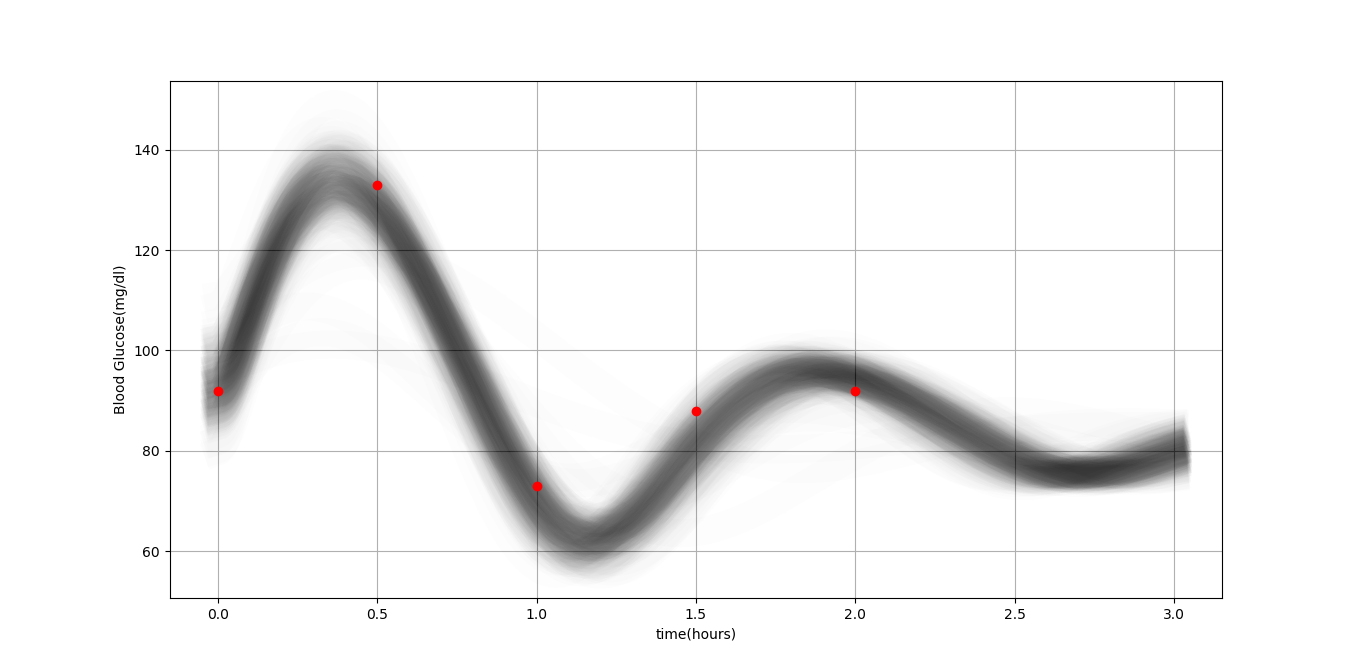}
    \end{tabular}
    \caption{OGTT inference for three patients. The first appears to be a healthy patient, the second a diabetic and the third an oscillating case. The graphs show the posterior distribution of $G(t)$ over 3 hours. Each vertical slice is a kernel density estimate of the posterior distribution of $G(t)$ at that time. The dots are the collected data.}
    \label{fig: realpatients}
\end{figure}

Figure \ref{fig: realpatients} shows a posterior sample for three real patients. The curves fit the data well, even for the third patient (top to bottom), whose data would not fit a curve which does not account for glucagon. The first patient is a healthy patient, whose body handles glucose normally. The second patient is a potentially diabetic patient, whose glucose does not return to the baseline during the test. The third patient is a patient whose body responds rapidly to glucose, causing oscillations. Performing inference on many patients has shown that the latter is not an unusual or rare situation.

0
These curves display more nuance than current guidelines or practices for OGTT analysis.  For instance, current practices would not distinguish between the first and third patients, despite their metabolism showing clearly different behavior, since the maximum measured value for both patients is similar (note also that although the first patient has higher glucose measurements, the third actually achieves a higher peak value in the estimated curve; this information can only be found by considering the temporal aspect of the measurements).  Our model has strong descriptive power, giving reasonably small uncertainty for times in the measurement interval. It also has reasonable predictive power for a short time outside of the measurement interval as can be seen by prolonging the function $G(t)$ beyond the last measurement (in our graphs we prolong this an additional hour.) This can be thought of as a projection of what the patient's glucose \textit{would} be if the conditions of the experiment were to continue. It is not clear, however, how long the dynamics of the system can be expected to remain intact, so this interpretation should only be considered over a short term.

\begin{figure}
    \begin{tabular}{ccc}
        \includegraphics[scale=0.075]{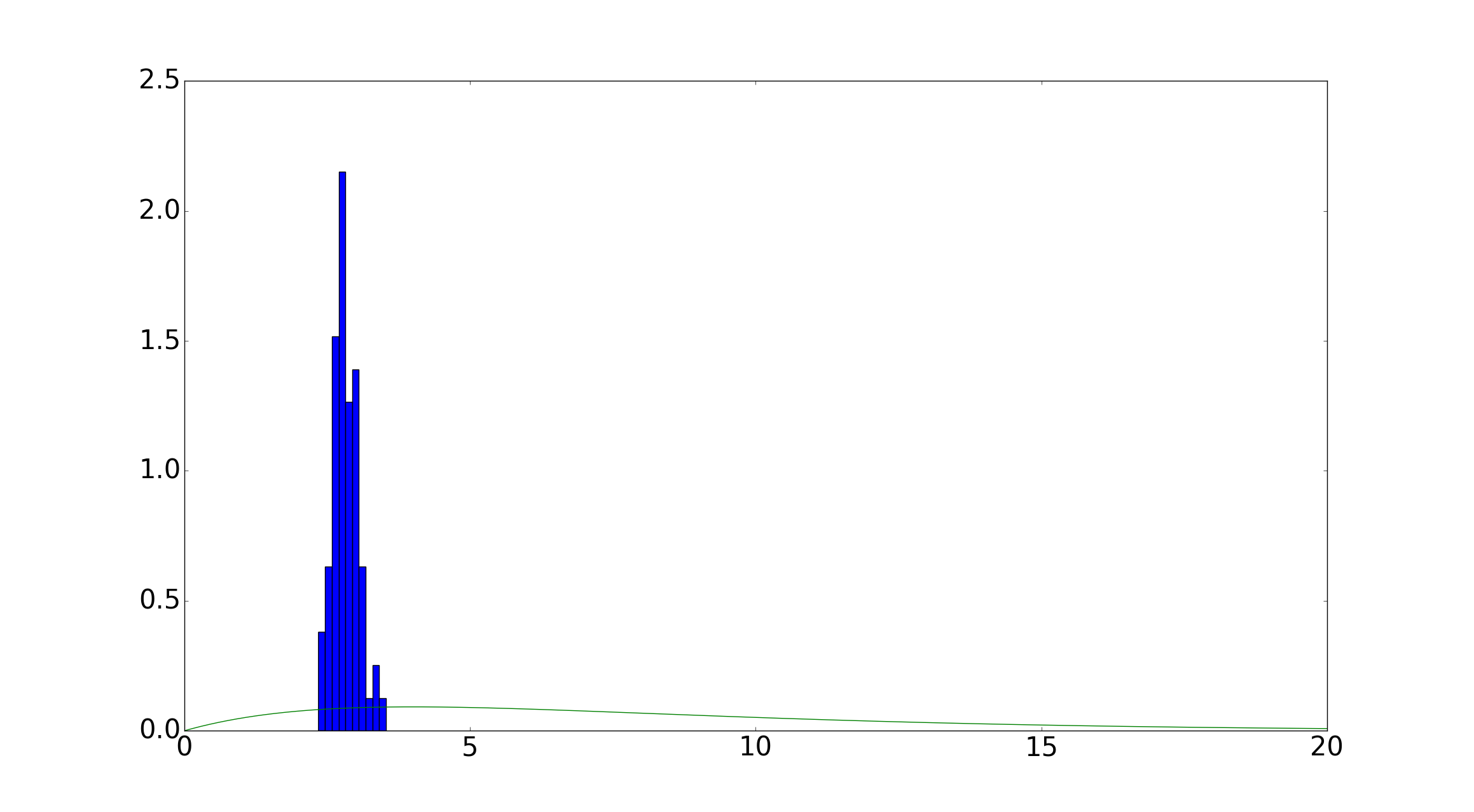} &
        \includegraphics[scale=0.075]{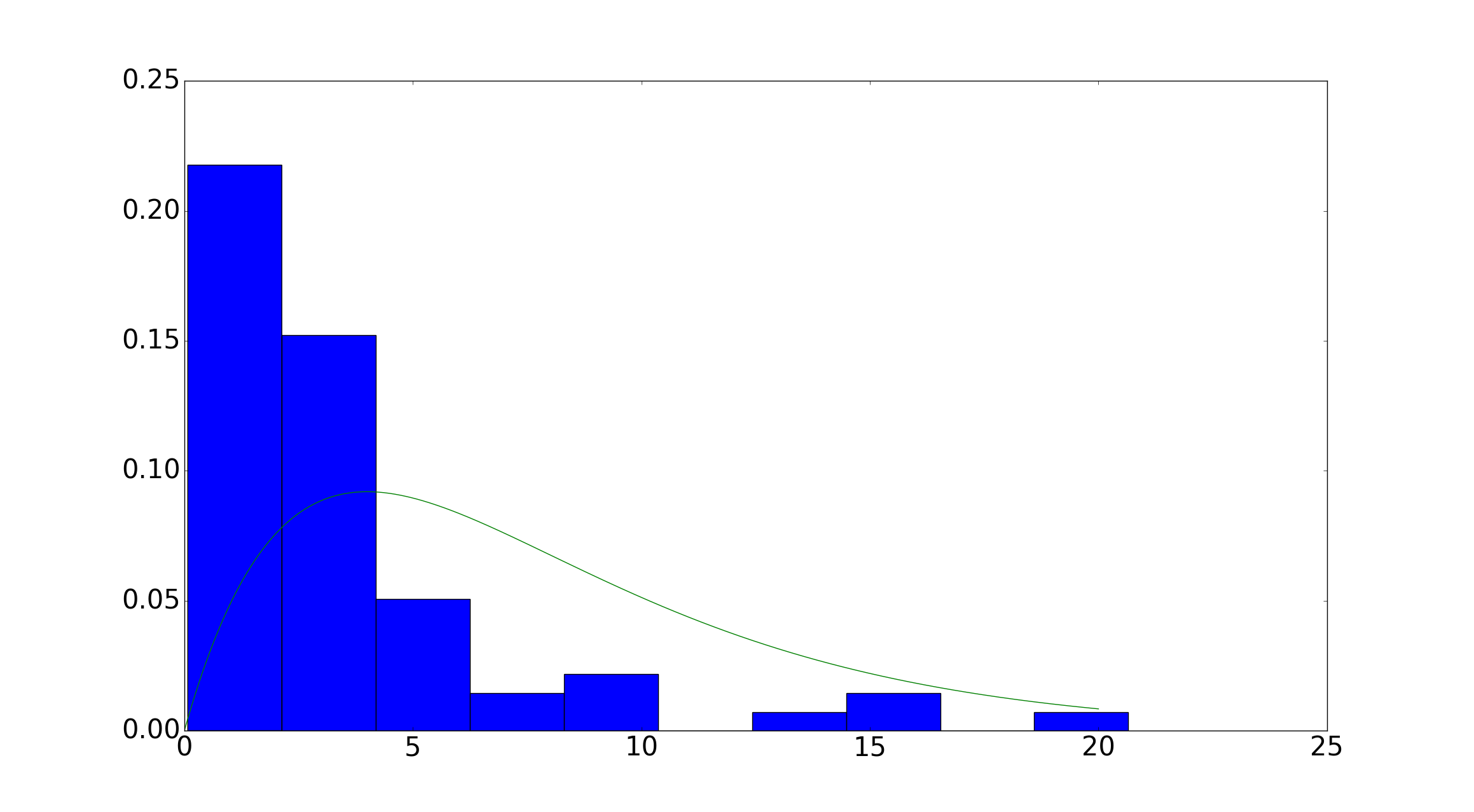} &
        \includegraphics[scale=0.075]{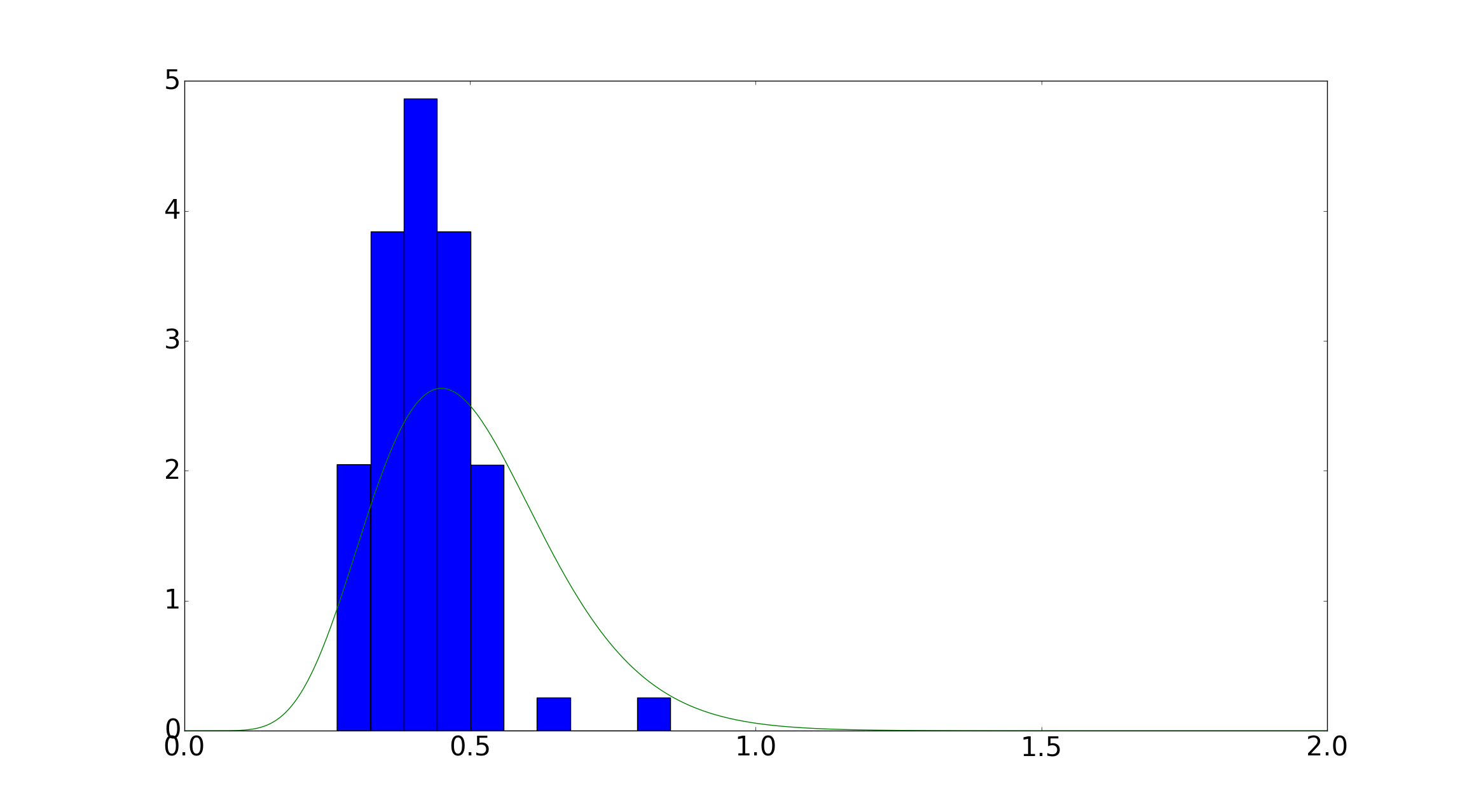} \\
        $\theta_0$ & $\theta_1$ & $\theta_2$
    \end{tabular}
        \caption{Histograms of posterior samples for the first patient.  They are the parameters $\theta_0$, $\theta_1$, and $\theta_2$ respectively.  They are superimposed on a graph of the prior density of each parameter.  In particular we note that for $\theta_1$ the posterior matches the prior closely, and for $\theta_0$, the data is extremely informative.}
\label{normalposteriors}
\end{figure}

\begin{figure}
    \begin{tabular}{ccc}
        \includegraphics[scale=0.075]{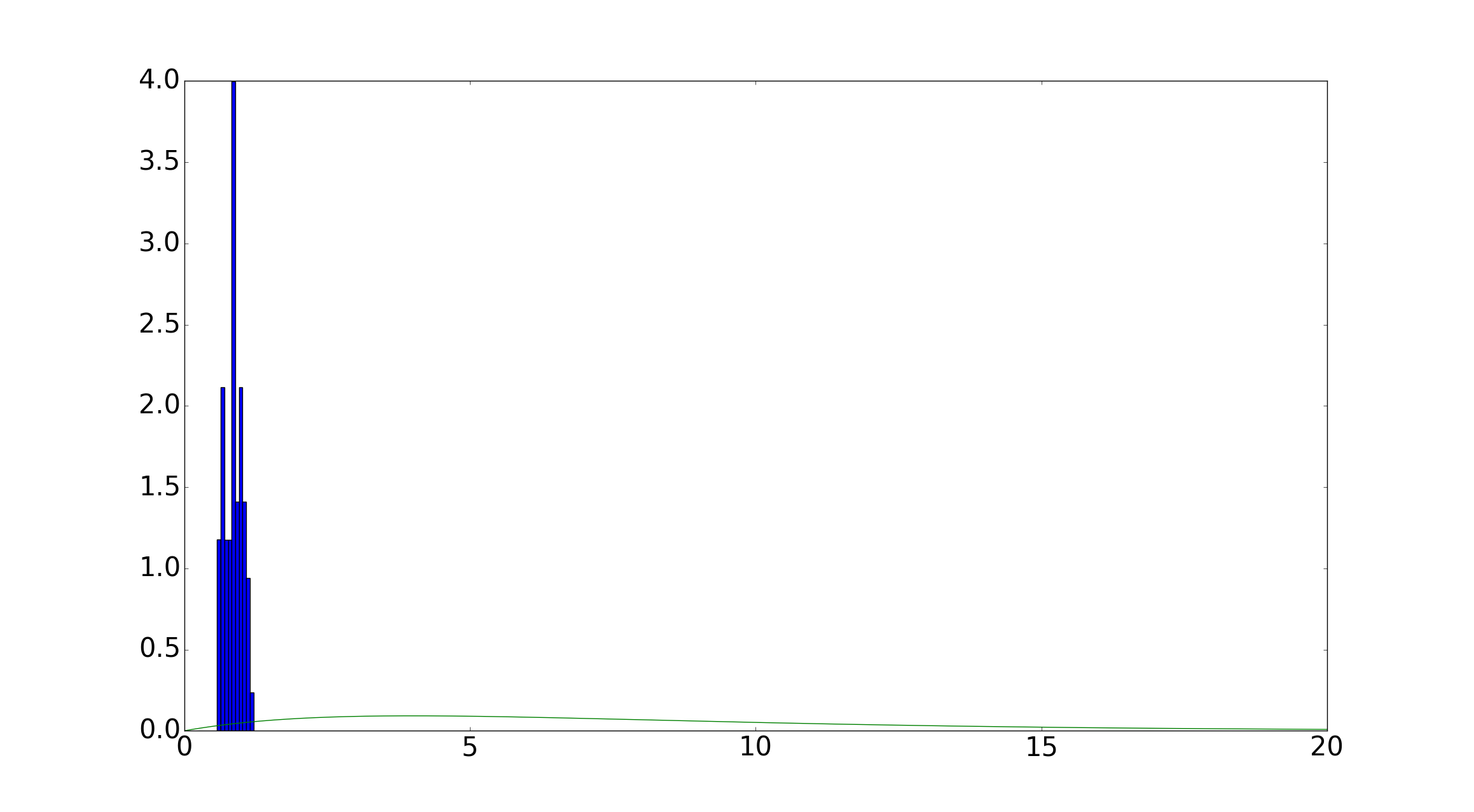} & \includegraphics[scale=0.075]{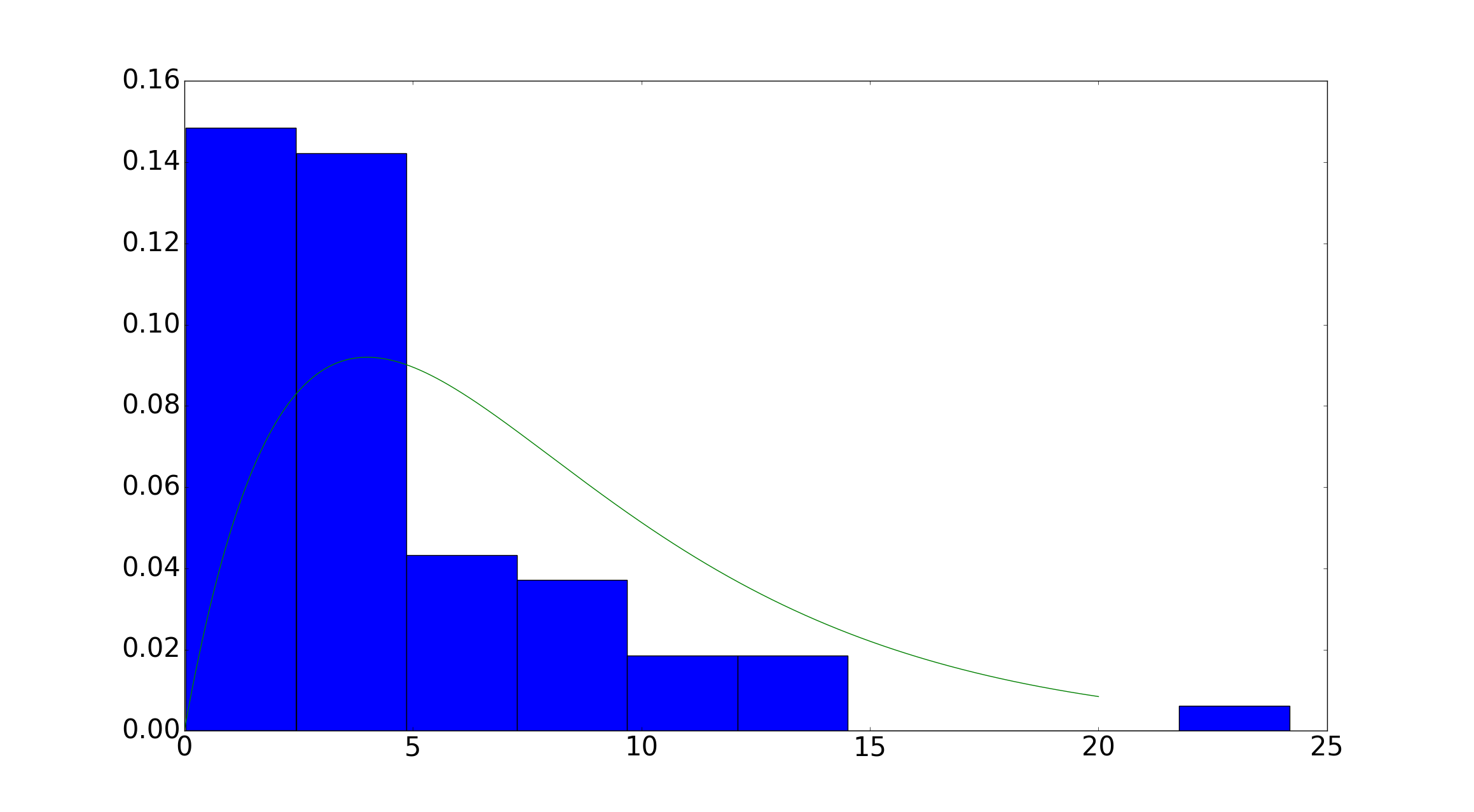} & \includegraphics[scale=0.075]{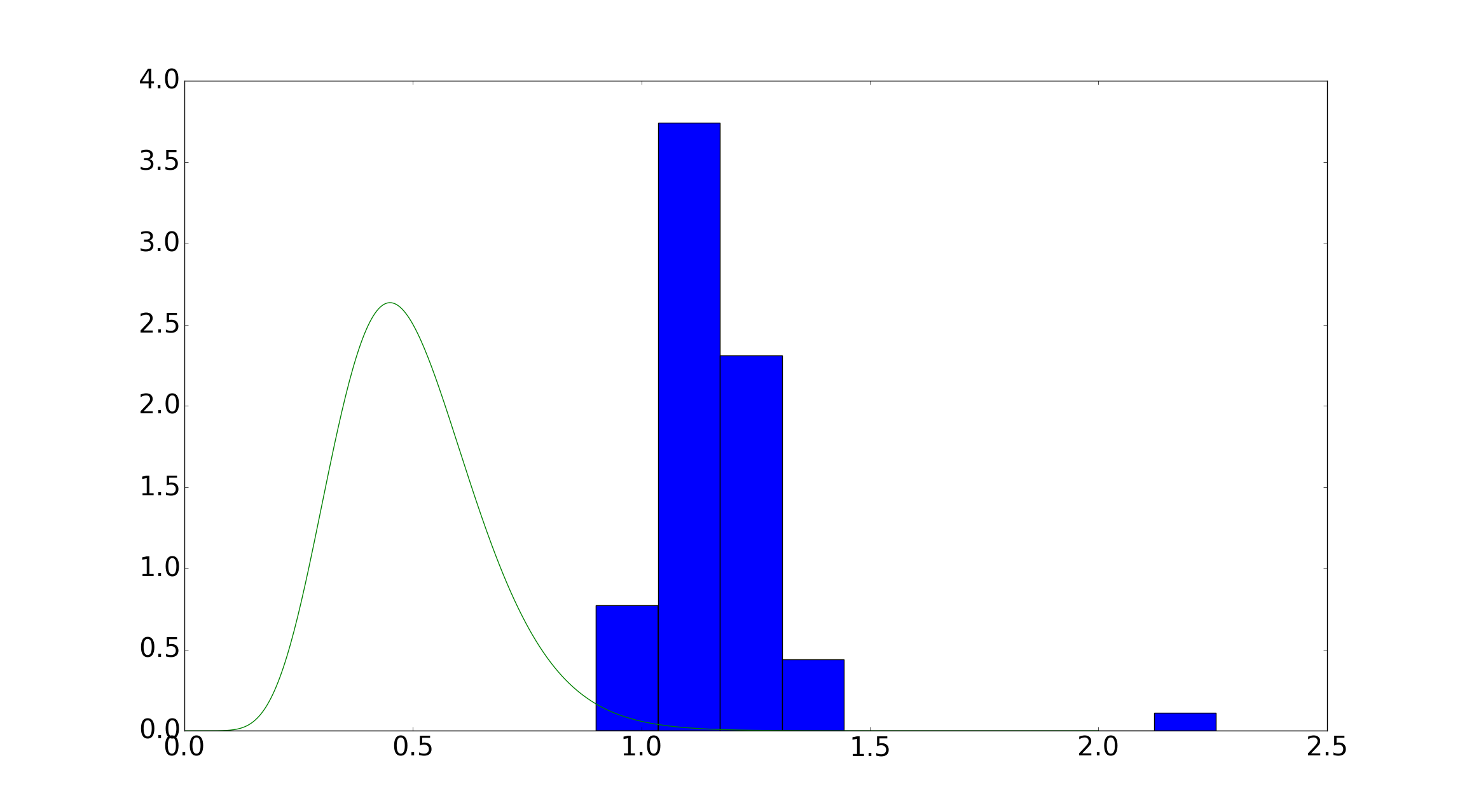} \\
        $\theta_0$ & $\theta_1$ & $\theta_2$
    \end{tabular}
    \caption{Similar graphs for the second (potentially diabetic) patient. Once again we note that $\theta_1$ once again deviates very little from the prior. This is caused by having no data below $G_b$.}
    \label{diabeticposteriors}
\end{figure}

\begin{figure}
    \begin{tabular}{ccc}
        \includegraphics[scale=0.075]{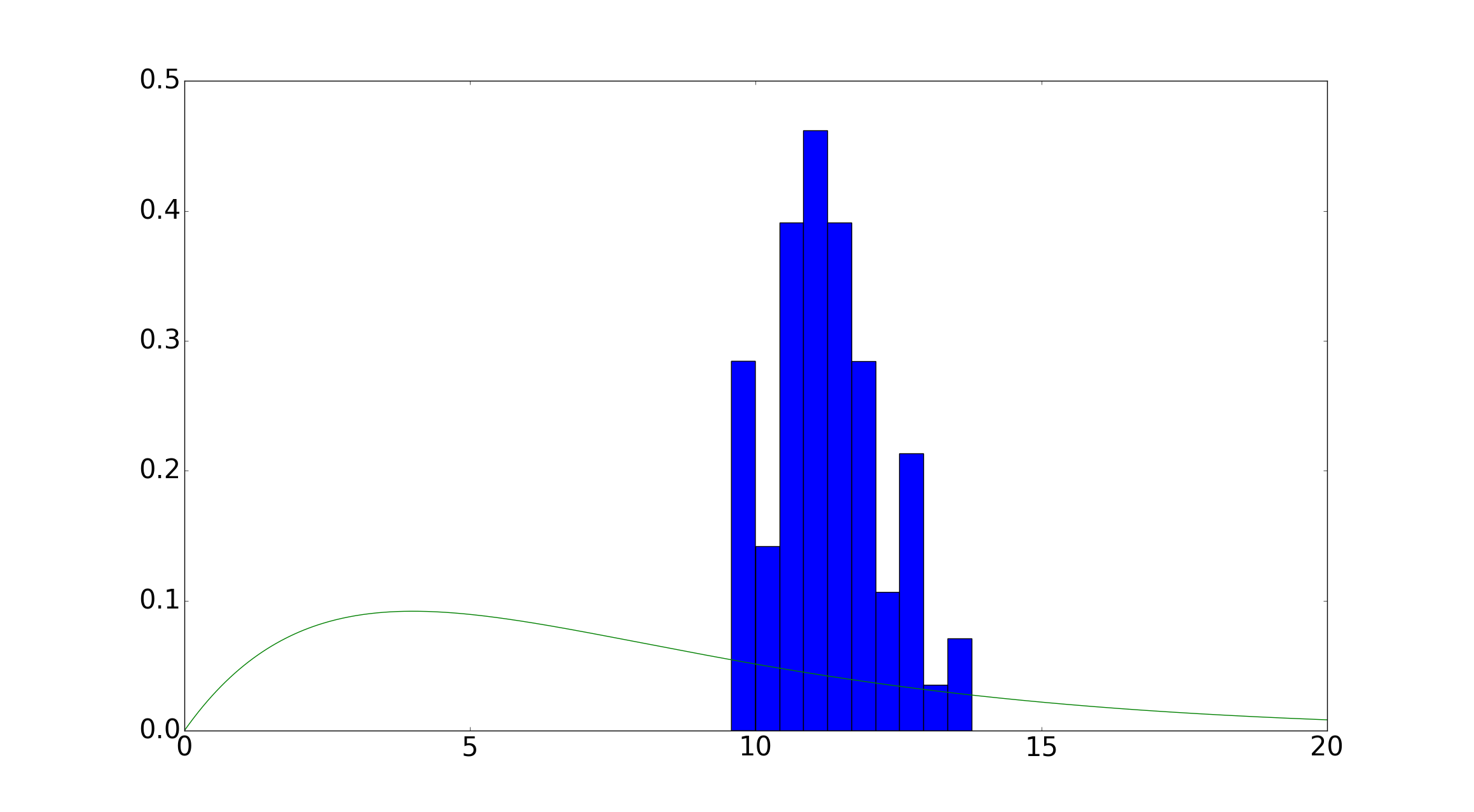} & \includegraphics[scale=0.075]{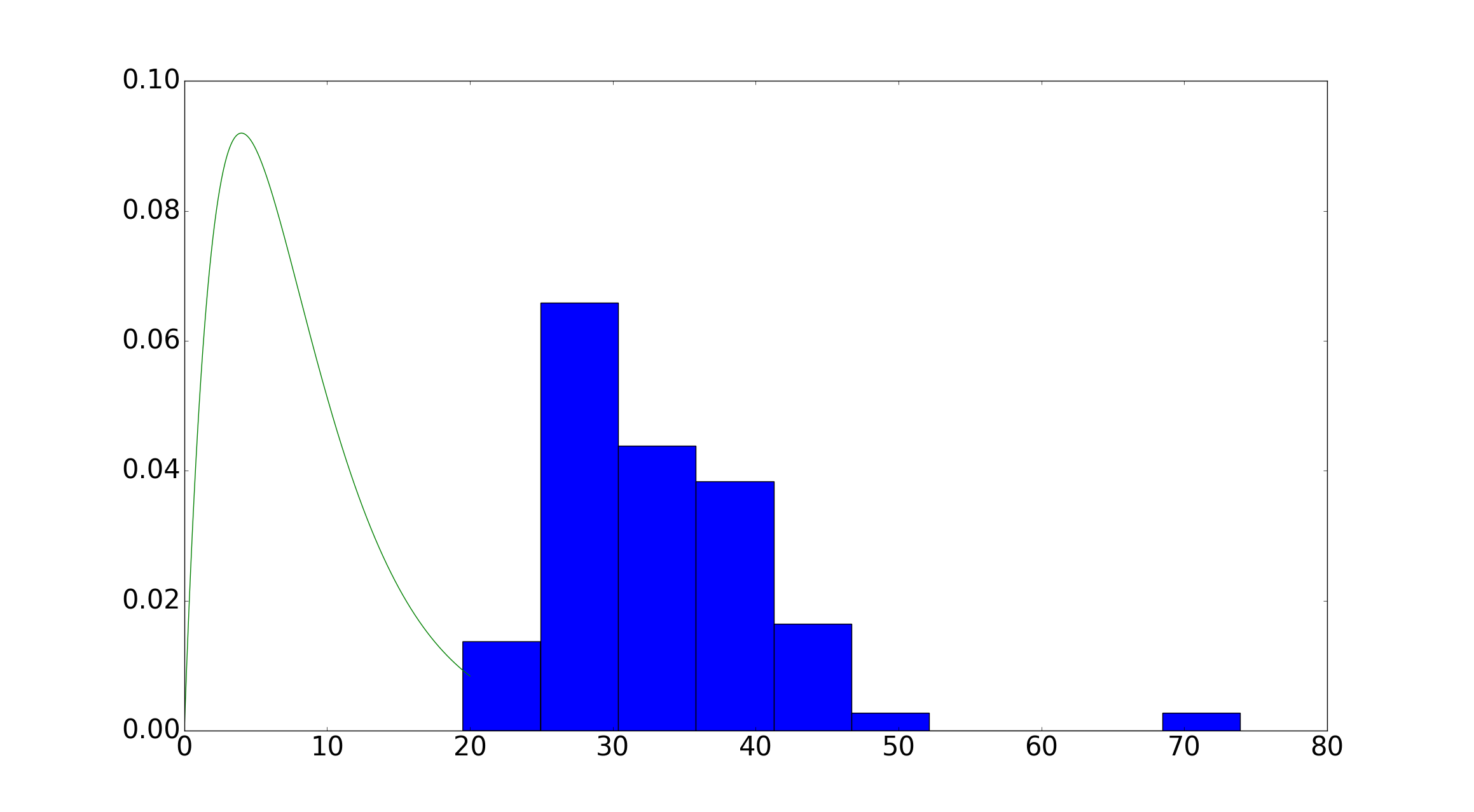} & \includegraphics[scale=0.075]{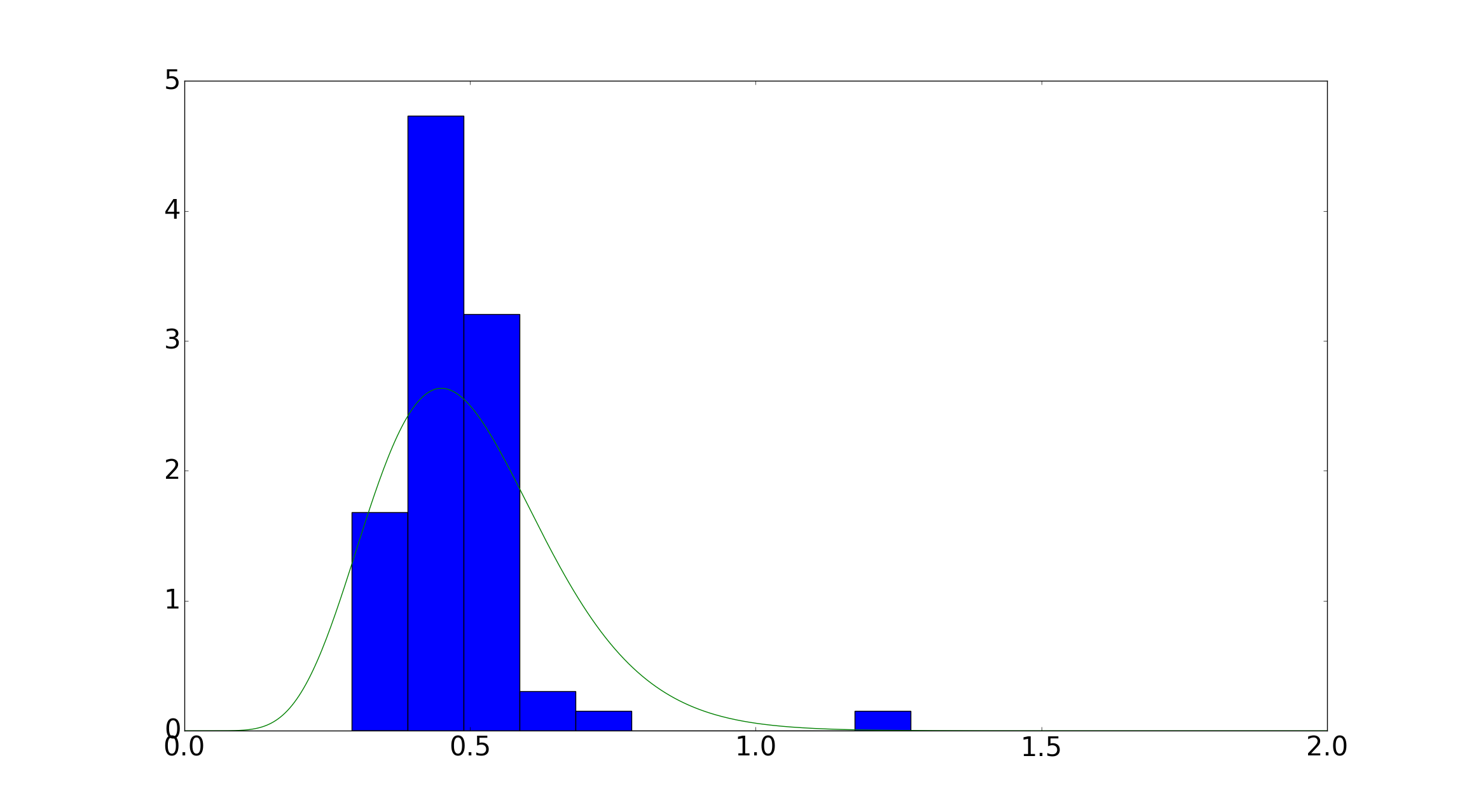} \\
        $\theta_0$ & $\theta_1$ & $\theta_2$
    \end{tabular}
    \caption{Similar graphs for the third (oscillating) patient. We note that in this case, the data that is below $G_b$ gives us information about $\theta_1$.}
    \label{oscillatingposteriors}
\end{figure}

Figures \ref{normalposteriors}, \ref{diabeticposteriors} and \ref{oscillatingposteriors} show histograms obtained from the MCMC posterior sampling for each of the model parameters for the patients from figure \ref{fig: realpatients}. The priors are represented with solid lines for reference. We may note that for patients without measurements below their resting glucose levels the data is uninformative about $\theta_1$, which represents glucagon response. This is to be expected since in our model glucagon does not kick in unless blood glucose goes below $G_b$. Oscillating patients do provide data that is informative with regards to $\theta_1$.

\section {Conclusions}\label{conclusions}

The diagnosis of type 2 diabetes is an important public health issue, and it requires a more sophisticated tool than the direct recording of values from the test, not only because these values are insufficiently informative, also because they do not account for measurement error.

Our model shows that overall it is able to represent the results of OGTT tests for nearly all patients for whom a fit was attempted. For one patient for whom the fit failed, it was later discovered that there was an error when recording the data, and the failure of the model to fit was an indication that triggered this error's discovery. The model also displays significantly deeper nuance and detail than previous analysis techniques could ever hope to represent.

At present, this model serves for the analysis of OGTT data, but not for diagnosis. The reason for this is that this model provides much more information than had previously been available, and our medical collaborators are - as of yet - uncertain about how to interpret this new information for which they have not been trained. Further study is required in order to transform full glucose curves into diagnoses or treatment recommendations. That said, this kind of a study should be well worth the effort.

At present, there is no known method which serves to diagnose initial stages of type 2 diabetes quickly, and accurate diagnosis may only be done by following a patient over time.  Regarding our model, we can envisage a faster and simpler solution based on a single dimensional marker. One single dimensional marker that seems reasonable is $min\left\{ t : G(t) =G_b ~\text{and}~ G'(t) < 0 \right\}$ (first return of blood glucose to the base level $G_b$), although simply using the marginal posterior distribution for
$\theta_0$, and comparing it with references $\theta_0$ values in healthy patients, might also be a possibility.  

We consider this model a strong candidate for further research in the analysis of OGTT data.  However,  even if not this specific model, some sort of dynamical model with strong descriptive power is required for the important and delicate tasks involved in the analysis of OGTT tests.

\section{Acknowledgements}

NK holds a CONACYT PhD grant.  NK and JAC are partially funded by CONACYT CB-2016-01-284451, RDECOMM and ONRG grants.

\section {Data Availability Statement}
The data that support the findings of this study are available from the corresponding author upon reasonable request.

\bibliography{esquema}

\begin{thebibliography}{}

\bibitem[\protect\citeauthoryear{Anderwald, Gastaldelli, Tura, Krebs,
  Promintzer-Schifferl, Kautzky-Willer, Stadler, DeFronzo, Pacini, and
  Bischof}{Anderwald et~al.}{2011}]{glucose}
Anderwald, C., A.~Gastaldelli, A.~Tura, M.~Krebs, M.~Promintzer-Schifferl,
  A.~Kautzky-Willer, M.~Stadler, R.~A. DeFronzo, G.~Pacini, and M.~G. Bischof
  (2011, Feb).
\newblock {{M}echanism and effects of glucose absorption during an oral glucose
  tolerance test among females and males}.
\newblock {\em J. Clin. Endocrinol. Metab.\/}~{\em 96\/}(2), 515--524.

\bibitem[\protect\citeauthoryear{Christen and Fox}{Christen and
  Fox}{2010}]{twalk}
Christen, J.~A. and C.~Fox (2010, June).
\newblock {A general purpose sampling algorithm for continuous distributions
  (the t -walk)}.
\newblock {\em Bayesian Analysis\/}~{\em 5\/}(2), 263--281.

\bibitem[\protect\citeauthoryear{Davidson, Schriger, Peters, and
  Lorber}{Davidson et~al.}{2000}]{ogtt}
Davidson, M.~B., D.~L. Schriger, A.~L. Peters, and B.~Lorber (2000, 08).
\newblock Revisiting the oral glucose tolerance test criterion for the
  diagnosis of diabetes.
\newblock {\em Journal of General Internal Medicine\/}~{\em 15}, 551--555.

\bibitem[\protect\citeauthoryear{Duckworth, Bennett, and Hamel}{Duckworth
  et~al.}{1998}]{insulindeg}
Duckworth, W.~C., R.~G. Bennett, and F.~G. Hamel (1998).
\newblock Insulin degradation: Progress and potential.
\newblock {\em Endocrine Reviews\/}~{\em 19\/}(5), 608--624.
\newblock PMID: 9793760.

\bibitem[\protect\citeauthoryear{Fox, Haario, and Christen}{Fox
  et~al.}{2013}]{inverse}
Fox, C., H.~Haario, and J.~Christen (2013).
\newblock Inverse problems.
\newblock In P.~Damien, P.~Dellaportas, N.~Polson, and D.~Stephens (Eds.), {\em
  Bayesian Theory and Applications}, Chapter~31, pp.\  619--643. Oxford
  University Press.

\bibitem[\protect\citeauthoryear{Jansson, Lindskog, Nordén, Carlström, and
  Scherstén}{Jansson et~al.}{1980}]{prev}
Jansson, L., L.~Lindskog, N.~Nordén, S.~Carlström, and B.~Scherstén (1980).
\newblock Diagnostic value of the oral glucose tolerance test evaluated with a
  mathematical model.
\newblock {\em Computers and Biomedical Research\/}~{\em 13}, 512--521.

\bibitem[\protect\citeauthoryear{Jiang and Zhang}{Jiang and
  Zhang}{2003}]{glucagon}
Jiang, G. and B.~B. Zhang (2003).
\newblock Glucagon and regulation of glucose metabolism.
\newblock {\em American Journal of Physiology-Endocrinology And
  Metabolism\/}~{\em 284\/}(4), E671--E678.

\bibitem[\protect\citeauthoryear{Jones, Oliphant, Peterson, et~al.}{Jones
  et~al.}{01  }]{python}
Jones, E., T.~Oliphant, P.~Peterson, et~al. (2001--).
\newblock {SciPy}: Open source scientific tools for {Python}.
\newblock [Online; accessed <today>].

\bibitem[\protect\citeauthoryear{Kaipio and Somersalo}{Kaipio and
  Somersalo}{2006}]{inversebook}
Kaipio, J. and E.~Somersalo (2006).
\newblock {\em Statistical and computational inverse problems}, Volume 160.
\newblock New York: Springer Science \& Business Media.

\bibitem[\protect\citeauthoryear{Metzger}{Metzger}{2006}]{diabetesinfo}
Metzger, B. (2006).
\newblock {\em American Medical Association guide to living with diabetes :
  preventing and treating type 2 diabetes : essential information you and your
  family need to know}.
\newblock Hoboken, N.J: John Wiley.

\bibitem[\protect\citeauthoryear{Palumbo, Ditlevsen, Bertuzzi, and
  Gaetano}{Palumbo et~al.}{2013}]{othermodel}
Palumbo, P., S.~Ditlevsen, A.~Bertuzzi, and A.~D. Gaetano (2013).
\newblock Mathematical modeling of the glucose–insulin system: A review.
\newblock {\em Mathematical Biosciences\/}~{\em 244\/}(2), 69 -- 81.

\bibitem[\protect\citeauthoryear{Petzold}{Petzold}{1983}]{lsoda}
Petzold, L. (1983).
\newblock Automatic selection of methods for solving stiff and nonstiff systems
  of ordinary differential equations.
\newblock {\em SIAM journal on scientific and statistical computing\/}~{\em
  4\/}(1), 136--148.

\bibitem[\protect\citeauthoryear{Wild, Roglic, Green, Sicree, and King}{Wild
  et~al.}{2004}]{prevalence}
Wild, S., G.~Roglic, A.~Green, R.~Sicree, and H.~King (2004).
\newblock Global prevalence of diabetes: Estimates for the year 2000 and
  projections for 2030.
\newblock \url{http://care.diabetesjournals.org/content/27/5/1047.full}.

\end{thebibliography}

\end{document}